\documentstyle[prl,aps,amsfonts,amssymb,floats,epsfig]{revtex}

\begin{document}

\draft
\twocolumn[\hsize\textwidth\columnwidth\hsize\csname @twocolumnfalse\endcsname

\title{Unifying Magnons and Triplons in Stripe-Ordered Cuprate Superconductors}

\author{G.S.~Uhrig,$^1$ K.P. Schmidt,$^1$ and M. Gr\"{u}ninger$^2$}
\address{$^1$Institut f\"{u}r Theoretische Physik, Universit\"{a}t zu
  K\"{o}ln, Z\"{u}lpicher Stra{\ss}e 77, D-50937 K\"{o}ln, Germany}
\address{$^2$II. Physikalisches Institut, Universit\"{a}t zu
  K\"{o}ln, Z\"{u}lpicher Stra{\ss}e 77, D-50937 K\"{o}ln, Germany}

\date{February 26, 2004; revised: July 21}

\maketitle

\begin{abstract}
Based on a two-dimensional model of coupled two-leg spin ladders, we derive a
unified picture of recent neutron scattering data of stripe-ordered
La$_{15/8}$Ba$_{1/8}$CuO$_4$, namely of the low-energy magnons around the
superstructure satellites and of the triplon excitations at higher energies.
The resonance peak at the antiferromagnetic wave vector ${\bf Q}_{\rm AF}$ in
the stripe-ordered phase corresponds to a saddle point in the dispersion of
the magnetic excitations. Quantitative agreement with the neutron data is
obtained for $J \! = \! 130-160$ meV and $J_{\rm cyc}/J \! = \! 0.2-0.25$.
\end{abstract}



\vskip2pc]
\narrowtext

Quantum magnetism in the cuprate superconductors is an intriguing issue.
A detailed understanding of the dynamic spin susceptibility
as measured by inelastic neutron scattering (INS) experiments should allow to
clarify the role of magnetism in the mechanism of high-$T_c$ superconductivity.
In particular two features have been in the focus of interest:
the appearance of the so-called resonance peak \cite{sidis04,norma03}
in the superconducting phase at the antiferromagnetic wave vector
${\bf Q}_{\rm AF}=(1/2,1/2)$  (see Fig.\ 1b)
at finite energies (e.g.\ 41 meV in optimally doped
YBa$_2$Cu$_3$O$_{7-\delta}$ (YBCO)) and the existence of stripe order which
manifests itself in superstructure satellites
around ${\bf Q}_{\rm AF}$ (Fig.\ 1b) \cite{norma03,tranq95,orens00}.
In general, these superstructure satellites are incommensurate, but they may
become commensurate for certain doping concentrations.
For many years, these two features have been regarded as separate issues,
each of them apparent in only one of the two families of cuprates on which most
neutron studies have focused: La$_{2-x}$Sr$_x$CuO$_4$ (LSCO) and YBCO.\@
But recent experimental results show that the resonance peak in YBCO
is accompanied at lower energies by incommensurate reflections
\cite{bourg00,rezni04},
and that stripe order appears also in YBCO \cite{mook02,mook98,arai99}.
Very recently, Tranquada {\em et al.} \cite{tranq04a} observed a resonance
peak at ${\bf Q}_{\rm AF}$
also in stripe-ordered La$_{15/8}$Ba$_{1/8}$CuO$_4$ (for $T>T_c$).

\begin{figure}[t]
\centerline{\psfig{figure=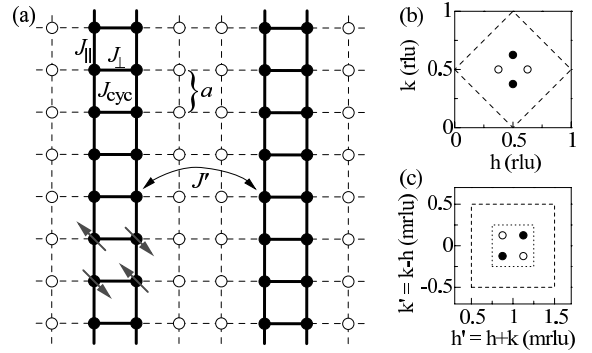,width=7.5cm,clip=}}
\caption{(a) 2D square lattice split into two-leg
$S$=1/2 ladders (full dots) coupled by
bond-centered charge stripes (open dots).
(b) Momentum space, coordinates in reciprocal lattice units $2\pi/a$ (rlu).
Dashed: magnetic Brillouin zone around the antiferromagnetic wave vector
${\bf Q}_{\rm AF}$=(1/2, 1/2). Open (full) dots: superstructure peaks from
vertical (horizontal) ladders.
(c) Momentum space rotated by 45$^\circ$ with $h^\prime \! = \! h+k$ and
$k^\prime \! = \! k-h$, thus ${\bf Q}^\prime_{\rm AF} \! = \! (1,0)$ in
magnetic reciprocal lattice units $2\pi/\sqrt{2}a$
(mrlu). Dotted: see caption of Fig.\ 4. } \label{fig:struktur}
\end{figure}

An $S$=1 collective mode (the resonance peak) in the superconducting phase
is a prominent feature of many different theoretical scenarios. Its
interpretation ranges from a particle-hole bound state (see Refs.\ in
\cite{sidis04,norma03}) to a  particle-particle bound state in SO(5) theory
\cite{demle98}.
In the stripe-ordered phase the choice of the
microscopic model is straightforward. Static stripe order
corresponds to a segregation into hole-rich charge stripes and hole-poor spin
ladders. Tranquada {\em et al.} \cite{tranq04a} analyzed their INS data at high
energies (including the resonance) in terms of the elementary triplet
excitations (triplons \cite{schmi03c}) of isolated {\em two}-leg ladders
(Fig.\ 1a) which are realized in case of bond-centered stripes \cite{anisi04}.
But the incommensurate low-energy excitations were described in a
separate model as spin waves (magnons), motivated by the existence of weak
long-range order.

Based on a model of coupled two-leg $S \! = \! 1/2$
ladders, we derive a unified description
of the low-energy superstructure modes, of the resonance peak
and of the high-energy excitations observed in Ref.~\cite{tranq04a}.
The superstructure modes arise from a ferromagnetic coupling between
neighboring ladders, whereas the resonance peak corresponds to a saddle point
of the triplon dispersion. The central result is that we arrive at a
quantitative agreement with the INS data for realistic values of the exchange
parameters.

The ladders shown in Fig.\ \ref{fig:struktur}a are
described by a standard Heisenberg Hamiltonian (see e.g.\ Ref.\
\onlinecite{schmi03d}) with rung coupling $J_\perp \! > \! 0$, leg coupling
$J_\parallel > 0$, and by the four-spin operators of a cyclic exchange
\begin{eqnarray}
H_{\rm cyc} &=& J_{\rm cyc}
\sum_i[({\bf S}_{i}^{\rm L}\cdot{\bf S}_{i}^{\rm R})
({\bf S}_{i+1}^{\rm L}\cdot{\bf S}_{i+1}^{\rm R}) + \\
&&\nonumber \hspace*{-8mm}
({\bf S}_{i}^{\rm L}\cdot{\bf S}_{i+1}^{\rm L})
({\bf S}_{i}^{\rm R}\cdot{\bf S}_{i+1}^{\rm R}) -
 ({\bf S}_{i}^{\rm L}\cdot{\bf S}_{i+1}^{\rm R})
({\bf S}_{i+1}^{\rm L}\cdot{\bf S}_{i}^{\rm R})
]\ ,
\end{eqnarray}
where $i$ counts the rungs, and $R$, $L$ label the two legs.
Inclusion of $J_{\rm cyc}\approx0.2-0.25 $
is crucial in order to obtain quantitative agreement
with experimental data, both in two-leg ladders \cite{nunne02} and in the 2D
cuprates \cite{colde01b,katan02}.
The ferromagnetic (FM) inter-ladder coupling $J'<0$, see
Fig.\ \ref{fig:struktur}a, is an effective coupling which
results from integrating out
the degrees of freedom in the stripes. If two spins on adjacent ladders
interact via a stripe rung with only one spin,
they prefer to be antiparallel to this spin, giving rise to an effective FM
coupling. The interaction via a half-filled stripe rung (two spins) implies an
antiferromagnetic (AFM) coupling. Plausibly, the more direct FM coupling
via one spin exceeds the AFM coupling via two spins,
so that a weak average FM coupling results.

\begin{figure}[t]
\centerline{\psfig{figure=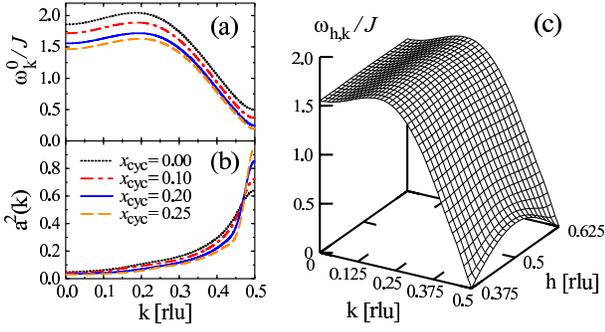,width=8cm,clip=}}
\caption{
(a), (b): dispersion $\omega^0_k$ and spectral weight $a^2(k)$
of one triplon in a two-leg $S$=1/2 ladder with $J_\parallel \! = \! J_\perp
\! = \! J$ and $x_{\rm cyc} \! = \! J_{\rm cyc}/J_\perp$.
(c) One-triplon dispersion $\omega_{h,k}$ for coupled ladders at the quantum
critical point $J^\prime/J \! = \! - 0.072$ for $x_{\rm cyc} \! = \! 0.2$.
}
\label{fig:input}
\end{figure}

We start from an effective model for isolated  ladders which we have
determined previously by perturbative CUTs (continuous unitary transformations)
\cite{schmi03d,knett01b}. The effective model is computed perturbatively to
order $n = 10$ which means that processes over a distance of up to 10 rungs
are taken into account. The one-triplon part is
$H_{\rm ladder} = \sum_{k,\alpha} \omega^0_k \; t^{\alpha,\dagger}_k \;
t^\alpha_k\ $,
where the operator $t^{\alpha,\dagger}_k$ ($t^{\alpha}_k$) creates
(annihilates) a hard-core boson -- a triplon \cite{schmi03c} -- with spin
$S=1$, momentum $k$ and one of the three triplet flavors $\alpha\in\{x,y,z\}$.
The dispersion $\omega^0_k$ (Fig.~\ref{fig:input}a)
is obtained as a series in $J_\parallel/J_\perp$ and
$x_{\rm cyc}=J_{\rm cyc}/J_\perp$ and evaluated by extrapolation
\cite{schmi03d,schmi03a}. To describe the square lattice we focus on
$J_\parallel = J_\perp$ where the extrapolation is reliable \cite{schmi03d}.
Inclusion of $J_{\rm cyc}$ reduces $\omega^0_k$ and in particular the gap at
$k\!=\! 1/2$ (by 50\% for $x_{\rm cyc}=0.2$).

We use the unitary transform of the observables $S_{i}^{\alpha,\text{R/L}}$
\begin{equation}
\label{obs-trafo1}
S_{i,\text{eff}}^{\alpha,\text{R}}
:= U^\dagger S_{i}^{\alpha,\text{R}} U
=
\sum_\delta a_\delta \, (t^{\alpha,\dagger}_{i+\delta} + t^\alpha_{i+\delta})
+ \ldots
\end{equation}
where the dots stand for normal-ordered quadratic
and higher terms in the bosonic operators. On the level linear in bosonic
operators, $S_{i,\text{eff}}^{\alpha,\text{L}} \!  = \!
-S_{i,\text{eff}}^{\alpha,\text{R}}$ holds
since rung triplets are odd excitations relative to rung singlets with respect
to reflection about the centerline of the ladder.
The Fourier transform of Eq.\ (\ref{obs-trafo1}) yields
\begin{equation}
S(k)_{\text{eff}}^{\alpha,\text{R}} =
a(k) (t^{\alpha,\dagger}_k + t^\alpha_{-k}) \; .
\label{eq:sk}
\end{equation}
Below we only need the one-triplon weight $a^2(k)$
(Fig.~\ref{fig:input}b), which is extrapolated in the same way as $\omega^0_k$.
Relative to the total weight $(S^\alpha_i)^2 \! = \! 1/4$ of the local
operator $S^\alpha_i$, the $k$-integrated one-triplon weight
$\int_0^1 a^2(k) \; dk$ is
74\% \cite{knett01b}, 70\%, 63\%, and 59\%
for $x_{\rm cyc}=$0, 0.1, 0.2, and 0.25. The missing weight belongs to
multi-triplon channels. By inclusion of $J_{\rm cyc}$, $a^2(k)$
becomes more pronounced around $k$=1/2.

The key idea is to couple neighboring ladders weakly by  $J'$ (see
Fig.~\ref{fig:struktur}a):
\begin{equation}
H' = -J'\sum_{h,k;\alpha}d_{h,k}
(t^{\alpha,\dagger}_{h,k}+t^{\alpha}_{-h,-k})(t^{\alpha}_{h,k}+
t^{\alpha,\dagger}_{-h,-k}) \; ,
\end{equation}
where $d_{h,k} \! := \! \cos(8\pi h) \, a^2(k)$ and $h$ is the  wave vector
perpendicular to the ladders measured in rlu of the square lattice,
cf.\ Fig.~\ref{fig:struktur}.
The total Hamiltonian $H$ is the sum of $H_{\rm ladder}$ for all ladders and
$H'$.
Since $J' \ll J$ it is justified to neglect the hard-core constraint so that
a Bogoliubov transformation yields
$H \! = \! \sum_{h,k;\alpha}\omega_{h,k} \, t^{\alpha,\dagger}_{h,k} \,
t^\alpha_{h,k}$ with
\begin{equation}
\omega_{h,k} \! = \! \sqrt{(\omega^0_k)^2-4J'd_{h,k}\omega^0_k} \; .
\label{eq:dispersion}
\end{equation}
In the INS data \cite{tranq04a} the minima are at $h \! = \! 1/2\pm 1/8$,
in agreement with a ferromagnetic $J'<0$.

We intend to describe the physics at the quantum critical point where magnons
just emerge \cite{sachd00}, i.e.\ we choose $J'$ such that
$\omega_{h,k} \! = \! 0$ at the minima. The corresponding
values for $J'$ are -0.20, -0.13, -0.072, and -0.051
for $x_{\rm cyc} \! = \! 0$,
0.1, 0.2, and 0.25. This is a physically plausible range \cite{dalos00}.
The overall shape of the 2D dispersion $\omega_{h,k}$
is depicted in Fig.\ \ref{fig:input}c. Magnons with linear dispersion emerge
from (1/2$\pm$1/8, 1/2). At ${\bf Q}_{\rm AF}$=(1/2, 1/2) a saddle point
occurs. The dispersion at higher energies is hardly changed from its form for
isolated spin ladders. This stems from the quadratic average
(\ref{eq:dispersion}) and from the suppression of the
one-triplon hopping for smaller values of $k$ due to the factor $a^2(k)$,
cf.\ Fig.\ \ref{fig:input}b.

\begin{figure}[t]
\centerline{\psfig{figure=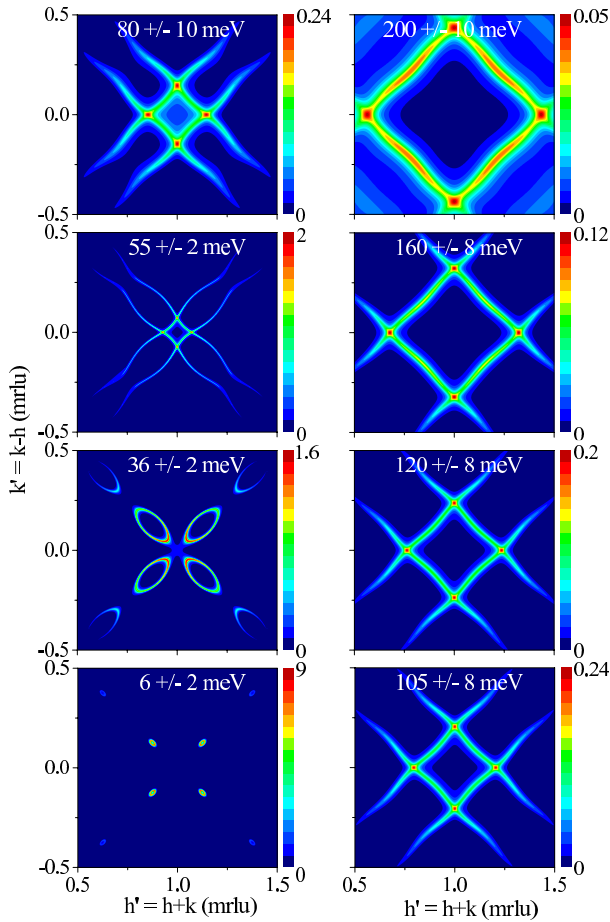,width=8cm,clip=}}
\caption{
Constant-energy slices of $\tilde{S}_{h^\prime,k^\prime}(\omega)$
for $J \! = \! 127$ meV, $J_{\rm cyc} \! = \! 0.2 J$ and
$J^\prime / J \! = \!  - 0.072$
with $\omega \pm \Delta \omega$
given in each panel for direct comparison with Fig.\ 2 in Ref.\
\protect\onlinecite{tranq04a}.
The experimental $\Delta \omega$ is accounted for by
$\omega\to\omega+i\Delta \omega$ in (\ref{eq:dynstruc}).
}
\label{fig:slices}
\end{figure}

The one-triplon part of the dynamic structure factor $S_{h,k}(\omega)$ reads
\begin{equation}
\label{eq:dynstruc}
S_{h,k}(\omega) =
-\frac{2}{\pi} \, {\rm Im}\frac{\sin^2(\pi h) \, a^2(k) \, \omega_k^0}
{(\omega+i0_+)^2-\omega^2_{h,k}}
\end{equation}
where the factor $a^2(k)$ stems from (\ref{eq:sk}), the factor $\sin^2(\pi h)$
from the interference of the two spins on a single rung, and $\omega_k^0$
from the Bogoliubov transformation. Both horizontal and vertical ladders
contribute to the INS data, summing over different stripe domains. Thus we
consider
$\tilde{S}_{h,k}(\omega) \! = (S_{h,k}(\omega) + S_{k,h}(\omega))/2$.
The constant-energy slices of $\tilde{S}_{h^\prime,k^\prime}(\omega)$ shown in
Fig.\ \ref{fig:slices} (in the rotated frame of Fig.\ 1c) agree very well with
the experimental results \cite{tranq04a}. At low energies (6 meV) the cones of
magnons emerging from the superstructure satellites
define a square around ${\bf Q}^\prime_{\rm AF}$.
The main intensity of these cones has shifted towards
${\bf Q}^\prime_{\rm AF}$ at 36 meV, culminating in the resonance peak at the
saddle point (44 meV). At still higher energies, the 1D character of the
triplons of a quantum spin liquid predominates, giving rise to almost straight
lines in the constant-energy slices. These lines form a diamond pattern which
is rotated relative to the low-energy square.

The high-energy diamond is filled in the INS data, but not in our single-mode
calculation. This discrepancy is caused
by the finite experimental momentum resolution,
by the decay of a single triplon (finite life time) into two or three
triplons due to the interladder coupling,
and by the missing two- and three-triplon channels of the isolated ladder (see above).
The two-triplon weight piles up around $k \! = \! 1/2$ \cite{knett01b},
which translates in 2D into a large contribution around ${\bf Q}_{\rm AF}$.
Energetically, the lower edge of the
three-triplon continuum is degenerate at ${\bf Q}_{\rm AF}$ with the
one-triplon energy (e.g.\ for triplons with $q_1\!=\!{\bf Q}_{\rm AF}$,
$q_2\!=\! - q_3 \! = \! (3/8,1/2)$, and $\omega(q_2)\!=\!\omega(q_3)\!=\!0$).
Thus both channels contribute to the filling of the diamond.

\begin{figure}[t]
\centerline{\psfig{figure=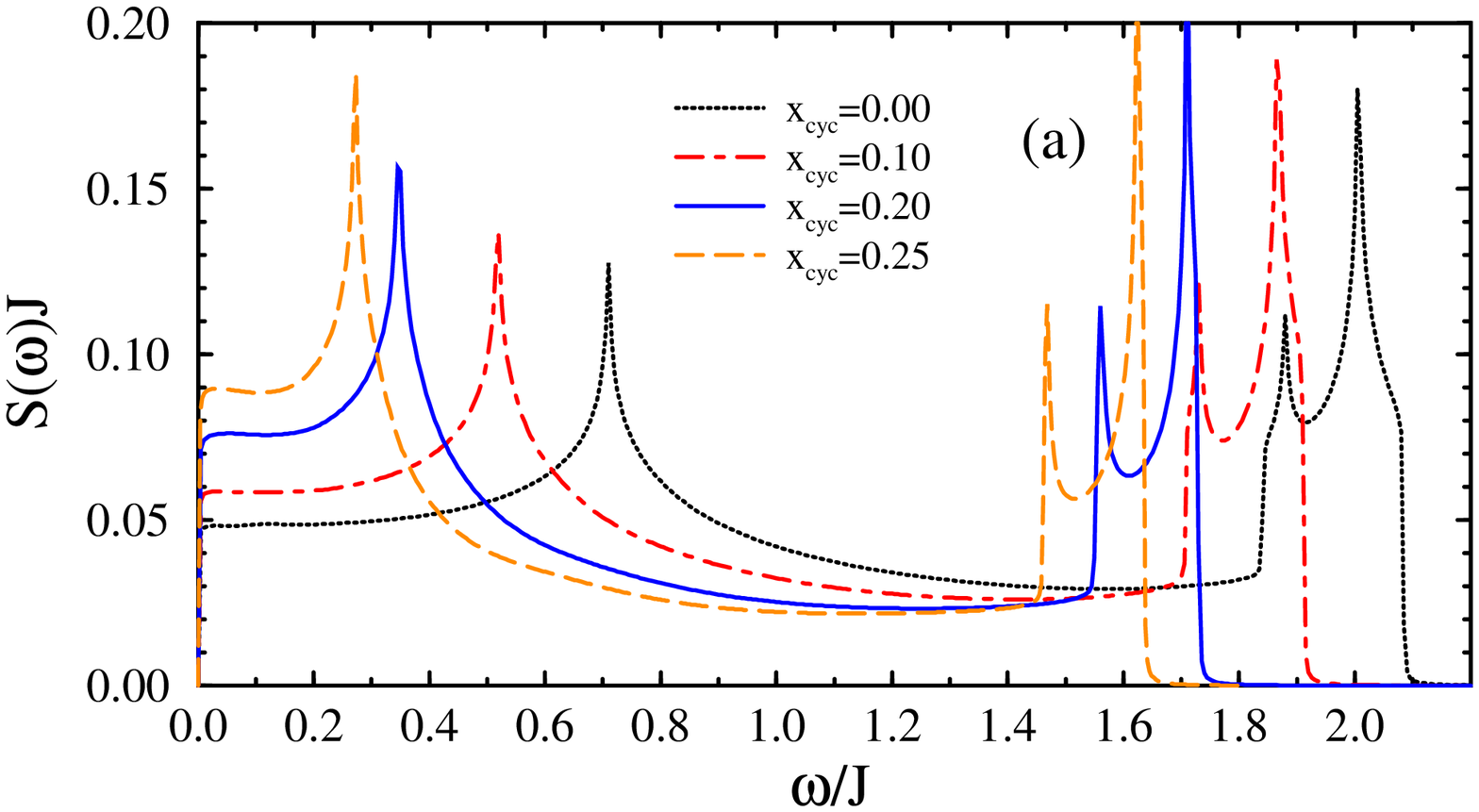,width=6.6cm,clip=}}
\centerline{\psfig{figure=uhrig_fig4b_rev.eps,width=6.6cm,clip=}}
\caption{
(a) Momentum-integrated structure factor $S(\omega)$.
(b) $S(\omega)$ integrated only over the magnetic Brillouin zone:
Symbols: INS data from Ref.\ \protect\cite{tranq04a}.
Lines:
For each value of $x_{\rm cyc} \! = \! 0$; 0.1; 0.2; 0.25
with $J^\prime/J$ chosen to meet the quantum critical point (see text),
we fix $J$ such that the lower singularity peaks at 44
meV ($J \! = \! 62$; 85; 127; 162 meV).
The intensity is not fitted but fixed by the value of $J$
(using a $g$-factor of $g \! = \! 2$).
The broadening $\Delta\omega$ is set to $2.5$meV.
Thin solid lines ($x_{\rm cyc} \! = \! 0.2$) mimic experimental details:
(i) at high energies, Lorentzian broadening $\Delta \omega \! = \! 20$meV;
(ii) at low energies: integration only in the dotted region of
Fig.\ \ref{fig:struktur}c \protect\cite{notiz1}.
}
\label{fig:Sw}
\end{figure}

For quantitative comparison we turn to the momentum integrated structure factor
$S(\omega)$. The one-triplon contribution is shown in Fig.\
\ref{fig:Sw}a. The saddle point gives rise to a logarithmic van Hove
singularity which can be identified with the resonance peak observed in Ref.\
\cite{tranq04a}. Qualitatively, pure spin wave calculations are similar in this
respect \cite{kruge03}.
A striking dependence of the singularity on $J_{\rm cyc}$ is
found. Values of $J \! \approx \! 140$-150 meV (for La$_2$CuO$_4$) and
$x_{\rm cyc} \! = \! 0.2$-0.3 are well established
\cite{nunne02,colde01b,katan02}.
Using $x_{\rm cyc} \! = \! 0.2$ (0.25), we find $J \! = \! 127$ (162) meV from
the INS resonance energy. This procedure fixes also the absolute scale of the
intensity (Fig.\ \ref{fig:Sw}b). The quantitative agreement
for the accepted exchange couplings strongly corroborates our approach.
The remaining differences can be explained easily.
(i) Our theory predicts a second sharp peak above 200 meV. There, the
experimental resolution is only $\pm 20-25$ meV, so that the high-energy peak
is washed out entirely (thin solid line in Fig.\ \ref{fig:Sw}b).
(ii) We find too much weight below $\approx $ 40 meV
because we do not consider a finite staggered magnetization.

The superstructure periodicity of $4 a$ with antiphase domain structure
manifests itself in the reciprocal space coordinates of the satellites
($n/8$, 1/2) and (1/2, $n/8$) with $n$ odd. This order has been depicted either
as {\em three}-leg ladders separated by a site-centered charge stripe or as the
bond-centered charge stripes discussed here and in Ref.\ \cite{tranq04a}.
Recent {\em ab initio} results \cite{anisi04} support the bond-centered
scenario. The quantitative agreement in Figs.\ 3 and 4 corroborates
the two-leg ladder scenario, but coupled three-leg ladders cannot be ruled
out at present.

Earlier, a unified description of the resonance peak and of the low-energy
satellites was proposed based on strictly 1D models \cite{batis01}. But in 1D
the satellites carry much more spectral weight than the resonance.
A true resonant behavior emerges only from a 2D saddle point.

In stripe-ordered $S=1$ La$_{2-x}$Sr$_{x}$NiO$_4$ (LNO)
\cite{bourg03,booth03}, isotropic magnons emerging from the superstructure
satellites describe the INS data up to the highest measured energy
($\approx 100$ meV) \cite{kruge03}. In the nickelates, a spin wave
picture is much more justified due to the larger spin $S\!=\!1$ and due to the
large interladder coupling $J^\prime$, which was found to be
$J^\prime/J \approx 0.5-1$ \cite{bourg03,booth03,kruge03}.
This shifts the saddle point, where the magnon cones merge, to high energies.
Indeed, enhanced intensity was observed at $\approx 80$ meV at
${\bf Q}_{\rm AF}$ \cite{bourg03}. In the cuprates, the saddle point is
rather low due to the small value of $J^\prime$ and
due to the sizeable value of $J_{\rm cyc}$ (Fig.\ 4a).

Our spin-only model aims at the magnetic response of the stripe-ordered phase.
Let us nevertheless speculate about the possible implications for the
magnetic properties of the {\em superconducting} phase. Since our calculation
does not rely on long-range magnetic order, it is plausible that its qualitative
features survive in the superconducting phase. The phase on the disordered side
of the quantum critical point shows a finite spin gap \cite{sachd00}, as
observed experimentally \cite{bourg00,rezni04,arai99}.
A saddle point exists if the dispersion is large in one direction and small in
the perpendicular direction. But in a metallic phase the damping due to
charge degrees of freedom has to be considered. In stripe-ordered
La$_{15/8}$Ba$_{1/8}$CuO$_4$ the spin-only model is applicable because the
damping is reduced by the charge order. Similarly, the appearance of the peak
and of the incommensurate low-energy modes below $T_c$ in nearly optimally
doped YBCO \cite{bourg00} can be attributed to the vanishing damping
if the resonance mode is lying below the particle-hole continuum.
Note the stunning similarity of Fig.\ 3 with recent INS data of YBCO
\cite{hayde04} showing depleted diamonds. The absence of the resonance peak
in LSCO, with similar $J$ but a smaller superconducting gap, suggests that
the mode is located {\em within} the continuum, where it is
overdamped \cite{morr98,abano99}.
The downward dispersion below the resonance energy $\omega_r$
\cite{bourg00,rezni04,mook02,arai99} as well as the upward dispersion
above $\omega_r$ \cite{rezni04,mook02,arai99} are generic features of our
stripe model. Close to $\omega_r$ the saddle point implies an almost circular
intensity distribution in $\tilde{S}_{h,k}$ which may reconcile recent INS
data of optimally doped YBCO \cite{rezni04} with a stripe scenario.

In conclusion, we have given a unified description of low-energy 2D magnons,
high-energy 1D triplons and of the resonance mode at ${\bf Q}_{\rm AF}$ in
stripe phases. Both the resonance energy and the spectral intensity
agree quantitatively with the INS data of La$_{15/8}$Ba$_{1/8}$CuO$_4$
\cite{tranq04a} for realistic values of $J\approx 130-160$ meV and
$J_{\rm cyc}/J \approx 0.2-0.25$. This underlines the significance of quasi-1D
quantum magnetism in the 2D cuprates.

We thank J.M. Tranquada for provision of the INS data and acknowledge fruitful
discussions with M. Braden and the support by the DFG via SFB 608 and SP 1073.

While we wrote this manuscript we became aware of a mean-field
calculation of $\omega_{h,k}$ based on coupled dimers \cite{vojta04}
leading to qualitatively similar results as in Fig.\ 3.


\end{document}